\documentclass[conference]{IEEEtran}
\IEEEoverridecommandlockouts

\usepackage{cite}
\usepackage{amsmath,amssymb,amsfonts}
\usepackage{algorithmic}
\usepackage{graphicx}
\usepackage{textcomp}
\usepackage{xcolor}
\usepackage[hyphens]{url}
\usepackage{enumitem}
\usepackage{hyperref}
\usepackage{caption}

\usepackage{setspace}
\begin{document}

\bstctlcite{IEEEexample:BSTcontrol}

\pdfpagewidth=8.5in
\pdfpageheight=11in

\pagenumbering{arabic}

\title{\huge Benchmarking Agentic HLS Design Tasks With HLS-Eval\vspace{-1ex}}
\author{
\IEEEauthorblockN{Stefan Abi-Karam$^{1,2}$, Callie Hao$^{1}$}
\IEEEauthorblockA{$^{1}$Georgia Institute of Technology, Atlanta, USA \quad $^{2}$Georgia Tech Research Institute, Atlanta, USA}
\IEEEauthorblockA{stefanabikaram@gatech.edu, callie.hao@gatech.edu}
}

\maketitle

\thispagestyle{plain}
\pagestyle{plain}

\begin{abstract}
Large language models (LLMs) and AI agents are increasingly explored for hardware design, including high-level digital design. While most work targets code generation and editing for hardware description languages (HDLs), our prior work introduced HLS-Eval, an open-source benchmark for evaluating LLMs on high-level synthesis (HLS) design tasks. Those evaluations, however, focused on zero-shot generation and editing, leaving open how agents achieve HLS design tasks.

We therefore extend HLS-Eval with an agentic evaluation flow built on the open-source mini-swe-agent framework. The flow lets HLS design agents use file-editing tools, invoke a C++ compiler for self-verification, and iteratively refine designs during inference, while logging agent traces for analysis of cost, token usage, and iteration count. We present initial results on the existing HLS-Eval benchmarks.

In our initial evaluation, we find open-source LLMs paired with an agentic harness solve every simple HLS code generation task in our evaluation, underscoring the need to expand benchmark difficulty as model capabilities advance. Analyzing traces from passing and failing runs, we show how model size, token usage, and trajectory length relate to design pass rates. These results establish a foundation for agentic HLS design and motivate harder benchmarks and new agentic tooling as model capabilities progress.
\end{abstract}

\section{Introduction}
Building domain-specific accelerators, such as for autonomous navigation and robotics \cite{Wan2022RoboticCO}, high-energy physics \cite{duarte2018fasta, kvapil2025intelligent}, and AI inference \cite{scalehls, allo, chen2024understanding} has traditionally required long design cycles and deep hardware expertise. FPGAs with high-level synthesis (HLS) help address this by enabling accelerators to be developed from high-level C, C++, or Python descriptions \cite{pyloga, scalehls, allo, dato}. By automating low-level tasks such as scheduling, binding, and dataflow generation, HLS promises to democratize domain-specific computing \cite{democratizing_dsc} by making hardware design more accessible to domain experts.

However, producing \textit{high-performance} HLS accelerators still requires substantial expertise. Designers must apply directives such as loop unrolling, array partitioning, and pipelining \cite{licht2020transformations}, structure dataflow and streaming computations, explore large parameterized design spaces, and reason about how source-level choices affect post-synthesis latency and resource use. This creates a steep barrier for non-experts, especially across diverse vendor and academic HLS toolchains.

Large language models (LLMs) have shown strong potential in software engineering \cite{llm_code} and HDL-based hardware design \cite{hdleval, verigen, verilogeval, verilogeval_v2, verilog_tool_feedback}, including code generation, optimization, and tool use. Yet comparable gains have not been realized for HLS \cite{hlseval}, where current methods remain far from the performance needed for practical domain-expert adoption. Although agentic LLM systems can plan, invoke tools, and solve multi-step tasks, their application to HLS workflows is still early \cite{agentichlsa, chathls, collini2025can, spec2rtlagent, synthai}.

Building on this observation, we identify a primary limitation that we propose to address: the field lacks comprehensive agentic HLS design benchmarks. Existing benchmarks fail to capture the full complexity of realistic HLS design tasks or quantify agent efficiency (e.g., cost and runtime) relative to achieved design performance. This gap in benchmarking infrastructure limits systematic evaluation across the hardware design research community, and we attribute it as a key reason progress in agentic HLS automation has lagged behind LLM-driven RTL design.

\begin{itemize}[leftmargin=*]
\item We extend our prior AI HLS design benchmarking work, HLS-Eval \cite{hlseval}, with an agentic evaluation flow built on a simple agent harness, enabling standardized assessment of LLM agents on HLS design tasks.
\item We evaluate this flow on HLS kernel generation from natural language descriptions and high-level specifications, reporting pass rates, inference scaling, and insights from agent traces.
\end{itemize}

\begin{figure*}[h!]
    \centering
    \includegraphics[width=0.8\linewidth]{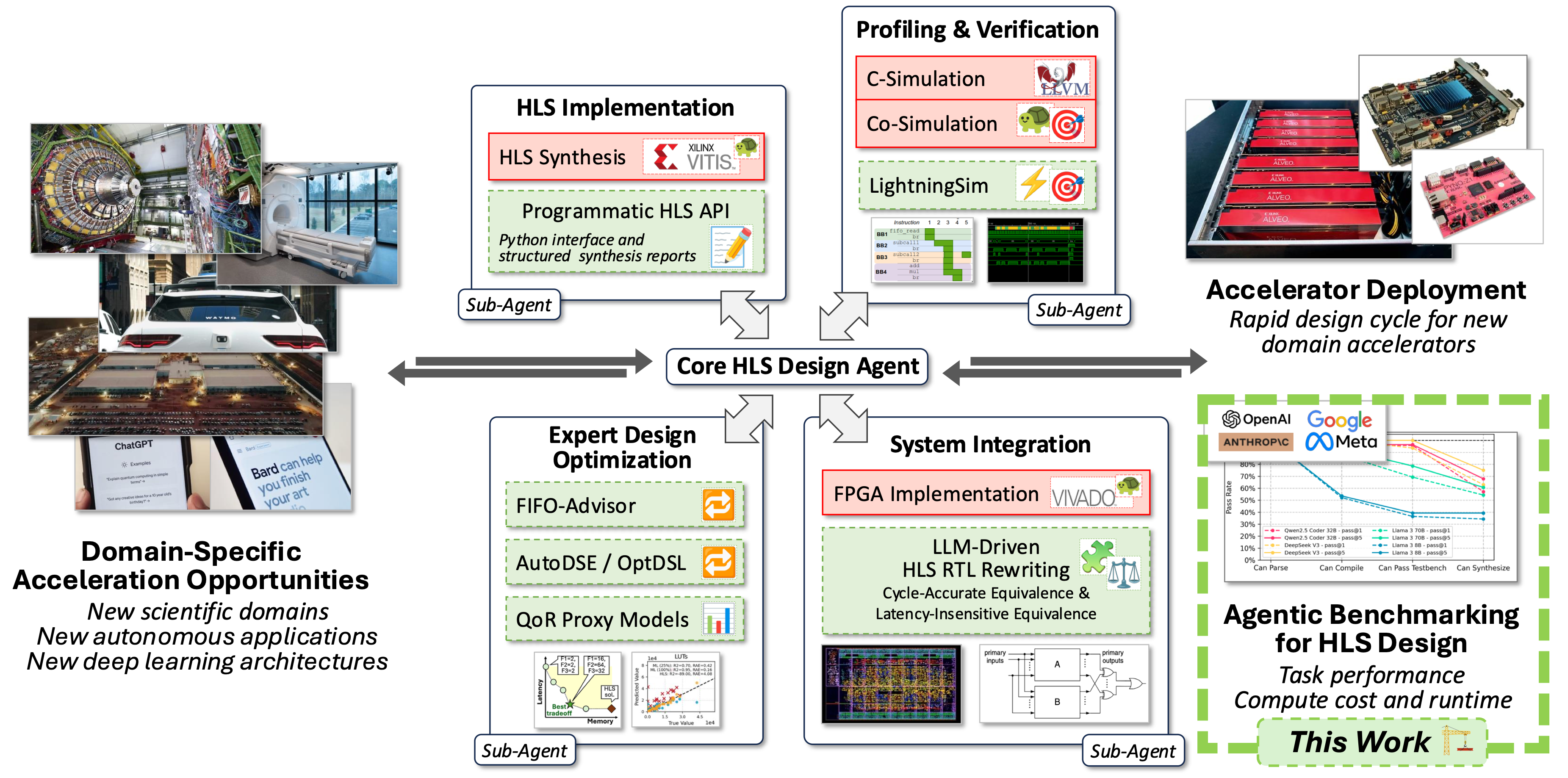}
\caption{Overview of our proposed end-to-end agentic HLS workflow for rapid prototyping of high-performance domain accelerators, with emphasis on extending HLS-Eval for benchmarking agentic systems for HLS design.}

    \label{fig:main_fig}
\end{figure*}

These results offer a starting point for integrating and benchmarking the proposed ideas into a larger, more capable agentic HLS design flow as highlighted in Figure \ref{fig:main_fig}.

\section{Methodology}
We augment HLS-Eval with a new evaluation flow built on the mini-swe-agent harness \cite{2026sweagent}, a simplified version of SWE-agent \cite{yang2024sweagent} widely used for reproducible LLM benchmarks on software engineering tasks. In this setup, the agent's only tool is a Bash shell inside a Linux Docker container, providing a minimal baseline with standard file utilities, C/C++ compilers, and scripting tools.

During evaluation, the agent receives an initial prompt and access to mounted design files within the container. After task completion or upon reaching the step or cost limits, we inspect the output directory for the final generated HLS implementation and verify that no provided testbench or header files were modified, preventing the agent from cheating on the evaluation.

We emphasize that this agentic flow directly subclasses the HLS-Eval \texttt{Evaluation} base class, and therefore inherits the rest of the HLS-Eval benchmarking infrastructure, including parallelization, automated HLS tool calls, and model and task parameterization. As a result, integrating the new flow, adapting our existing run scripts, and collecting our new results took less than one PhD-student hour.

Complete details of the implementation can be found in the HLS-Eval open-source code repository: \texttt{\href{https://github.com/sharc-lab/hls-eval}{https://github.com/sharc-lab/hls-eval}}.

\section{Results}
We evaluate our proposed agentic evaluation flow on the set of HLS designs already present within HLS-Eval: a total of 85 designs from the CHStone \cite{chstone}, MachSuite \cite{machsuite}, Polybench \cite{polybench}, Rosetta \cite{rosetta}, and C2HLSC \cite{c2hlscb} design sources. We chose to evaluate the \texttt{gpt-oss-20b} and \texttt{gpt-oss-120b} LLM models within our agentic harness. We selected these models for our initial evaluation because they represent strong, low-cost, open-source baselines with two model size variations for comparison.

\subsection{HLS Design Generation Task}

\begin{figure}
    \centering
    \includegraphics[width=\linewidth]{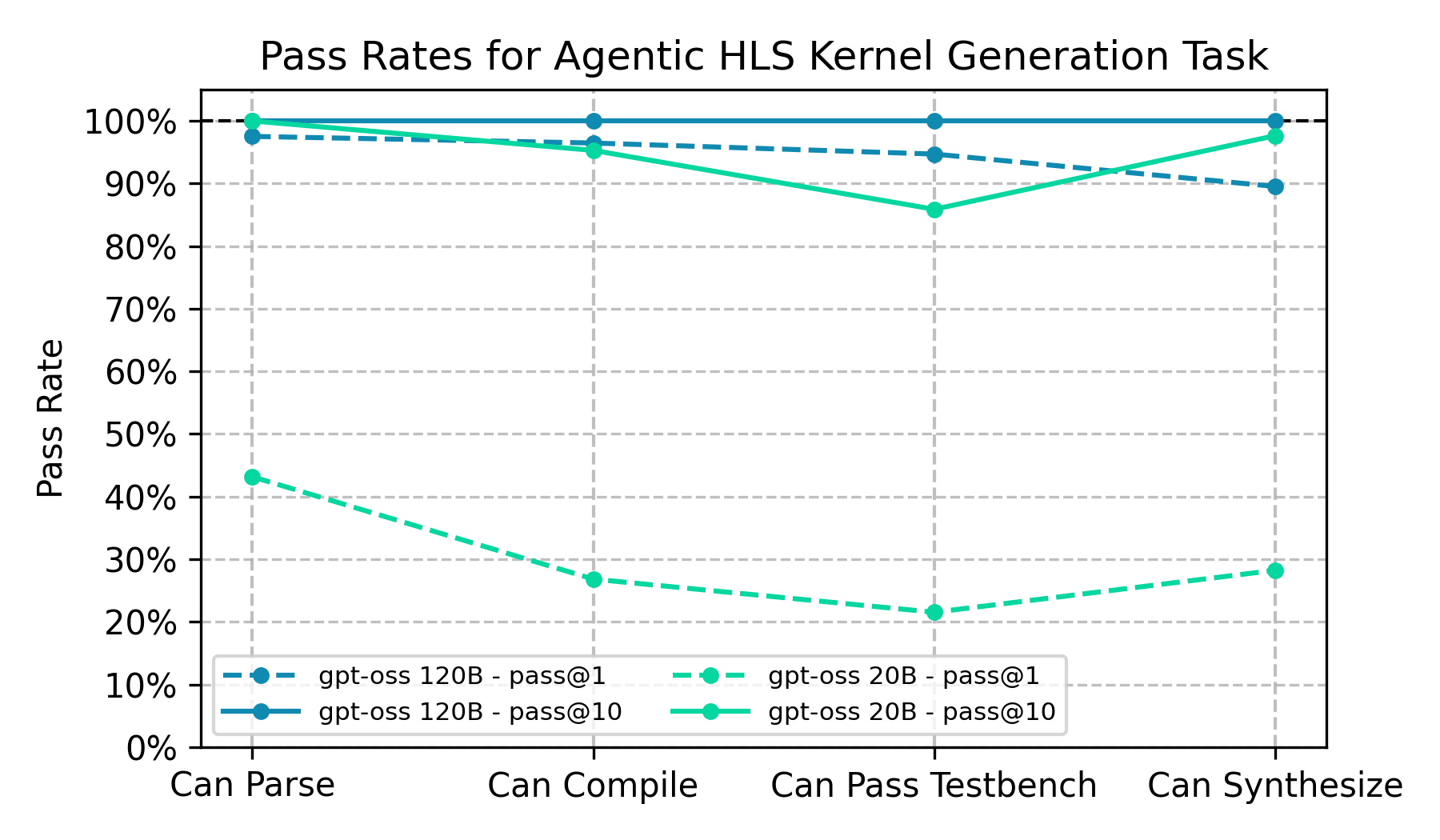}
    \caption{Pass rates for mini-swe-agent at each stage of the HLS design flow. The \texttt{gpt-oss-120b} model saturates the benchmark with a $100\%$ pass rate at all stages with $k=10$.}
    \label{fig:pass_rates}
\end{figure}

We present pass rate results for the HLS design generation task in Figure~\ref{fig:pass_rates}. For each model, we evaluate the pass@k rate across multiple stages of design validation: parsing the design from the agent's output, checking whether the generated design compiles, checking whether the compiled design passes a functional testbench, and checking whether the HLS tool successfully synthesizes the design. We report pass@k for k=1 and k=10, with N=10 samples per evaluation case.

Most strikingly, \texttt{gpt-oss-120b} completely saturates the benchmark at the pass@10 rate. Its corresponding pass@1 rate also exceeds $90\%$ for all HLS design stages. This shows that even a modestly sized open-source model, relative to larger commercial and open-source models, can generate simple kernels that are both synthesizable and correct, given only a natural-language specification and a testbench harness. We attribute this capability to the model's ability to call a C++ compiler for syntactic and functional self-verification at inference time, which highlights the key advantage of agents over naive zero-shot inference.

As a result, since saturated results provide no further informative feedback to measure AI models' HLS design capabilities, they motivate the development of more complex benchmark design cases that represent tasks akin to developers building full end-to-end domain accelerators.

\subsection{Verifiers and Inference Scaling}
\label{sec:scaling}

\begin{figure}
    \centering
    \includegraphics[width=0.75\linewidth]{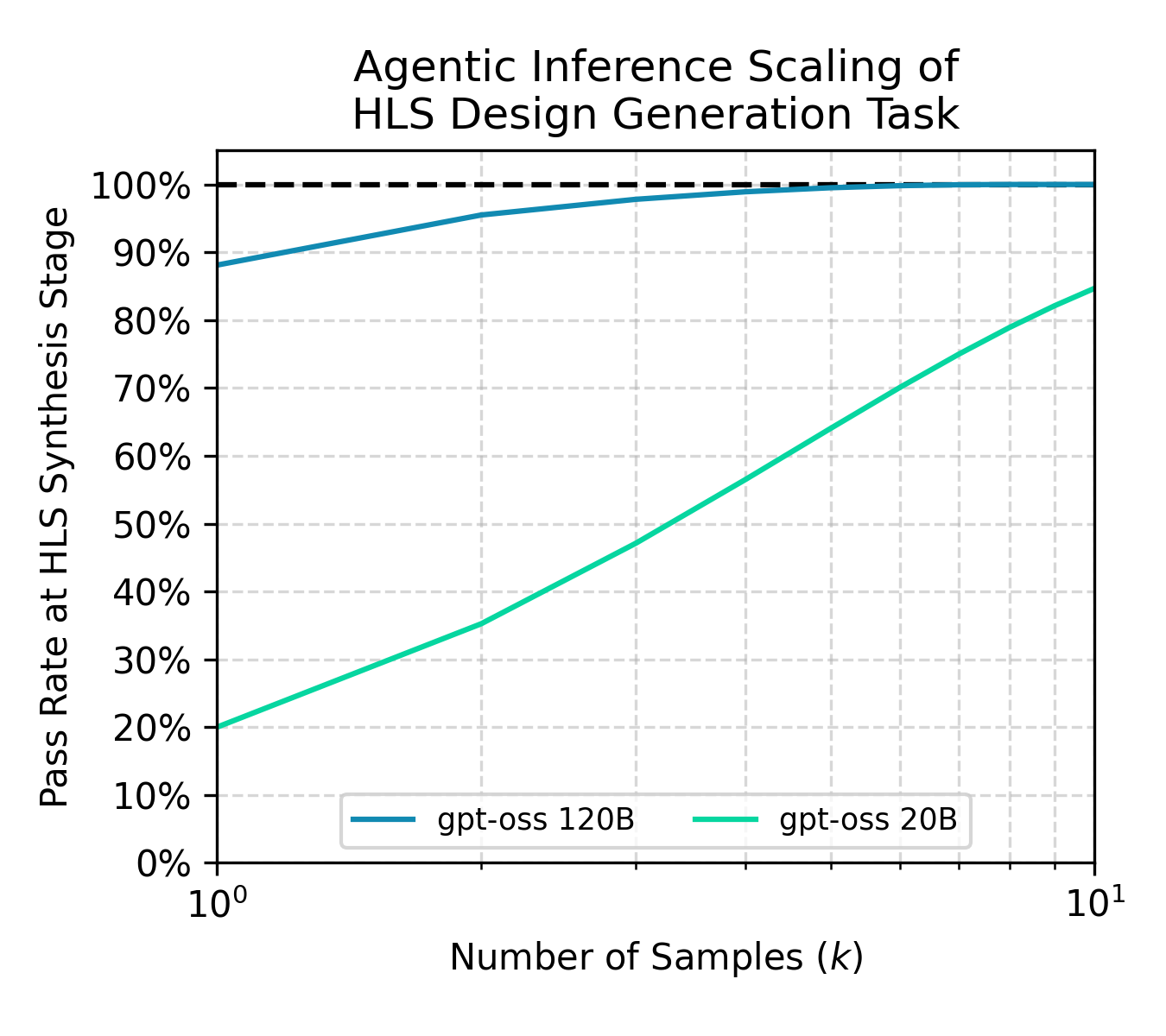}
    \vspace{-0.5em}
    \caption{Inference scaling results on pass rate for agentic HLS kernel generation task. The smaller model can recover $+60\%$ pass rate with $k=10$ samples.}
    \label{fig:inference_scaling}
\end{figure}

Many stages of the HLS design flow act as strong verifiers for LLM-based agents, enabling self-checking of design correctness, functionality, and performance during inference. C++ compilers, co-simulation tools, HLS synthesis tools, design space exploration (DSE) frameworks, and downstream implementation tools serve as verifiers when accessible to the agent, and also enable inference scaling. Since only one agent rollout needs to produce a design that passes all verification stages, sampling more agent rollouts for the same task can substantially improve pass rates \cite{brown2024large} and overall design quality.

As shown in Figure~\ref{fig:inference_scaling}, our agentic evaluations exhibit this inference scaling behavior. We observe large improvements from $k=1$ samples to $k=10$ samples, with gains depending on model size. The effect is strongest for weaker models: the \texttt{gpt-oss-20b} model achieves a $+60\%$ increase in synthesis pass rate at $k=10$, while the larger \texttt{gpt-oss-120b} model sees a $+11\%$ improvement.

The sampling strategy described represents only one form of inference scaling. Future work will explore structured LLM-assisted evolutionary search \cite{alphaevolve} and methods that balance inference scaling benefits with model inference cost and tool runtime overhead.

\subsection{Agent Trajectories}

\begin{figure}[t]
    \centering
    \includegraphics[width=\linewidth]{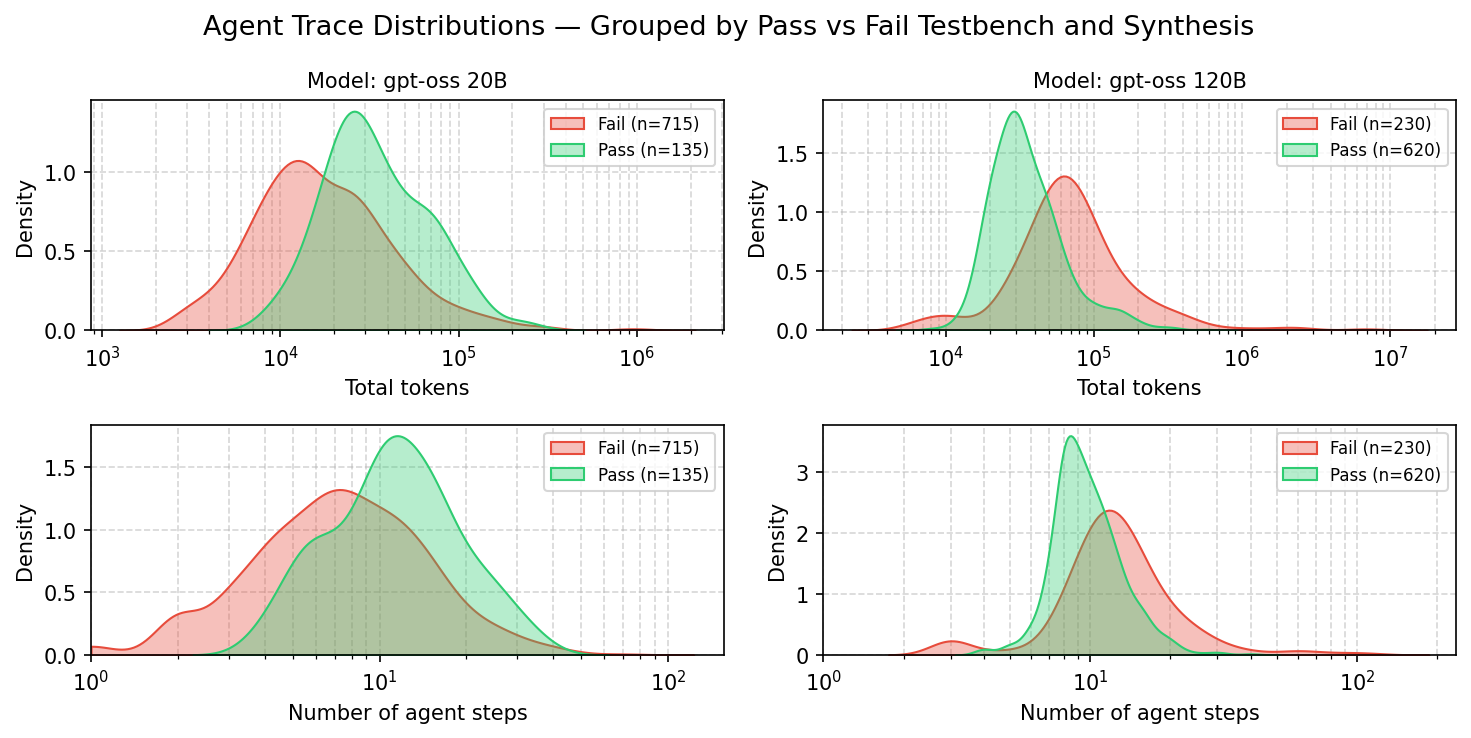}
    \caption{Distribution of token counts and trajectory lengths for agent runs, grouped by model and evaluation outcome.}
    \label{fig:agent_trace}
\end{figure}

To better understand agent behavior on HLS design tasks, we present an initial analysis of inference traces. Figure~\ref{fig:agent_trace} shows the distribution of token counts and trajectory steps, grouped by whether the design generated during a trace passed or failed the testbench and synthesis. We observe model-dependent trends: for the smaller \texttt{gpt-oss-20b} model, passing traces use more tokens and more steps, whereas for the larger \texttt{gpt-oss-120b} model, failing traces use more tokens and more steps.

Although the causes of these dynamics and their relationship to model scale and task complexity remain unclear, these findings motivate deeper investigation. Future work will incorporate more HLS-specific trace analysis, including tool usage frequency and correlations between tool runtimes and final design performance.

\section{Conclusion}

We present our initial effort to extend HLS-Eval toward end-to-end agentic design of domain-specific accelerators using HLS. We report preliminary results on open-ended agentic evaluation for HLS design tasks, and we explore inference scaling and trace analysis in the context of agentic HLS design. By combining LLM-driven HLS agents, advanced tool automation, and standardized benchmarking within reproducible open-source frameworks, we aim to enable scalable, intelligent, and accessible domain acceleration for scientists, engineers, and researchers alike.

\bibliographystyle{IEEEtranS}
\bibliography{refs}

@IEEEtranBSTCTL{IEEEexample:BSTcontrol,
CTLuse_forced_etal       = "yes",
CTLmax_names_forced_etal = "3",
CTLnames_show_etal       = "2" }

@inproceedings{hlseval,
	location = {Stanford, {CA}, {USA}},
	title = {{HLS}-Eval: A Benchmark and Framework for Evaluating {LLMs} on High-Level Synthesis Design Tasks},
	rights = {https://doi.org/10.15223/policy-029},
	isbn = {979-8-3315-2597-2},
	url = {https://ieeexplore.ieee.org/document/11106033/},
	doi = {10.1109/ICLAD65226.2025.00021},
	shorttitle = {{HLS}-Eval},
	eventtitle = {2025 {IEEE} International Conference on {LLM}-Aided Design ({ICLAD})},
	pages = {219--226},
	booktitle = {2025 {IEEE} International Conference on {LLM}-Aided Design ({ICLAD})},
	publisher = {{IEEE}},
	author = {Abi-Karam, Stefan and Hao, Cong},
	urldate = {2026-03-24},
	date = {2025-06-26},
}

@online{brown2024large,
  title = {Large {{Language Monkeys}}: {{Scaling Inference Compute}} with {{Repeated Sampling}}},
  shorttitle = {Large {{Language Monkeys}}},
  author = {Brown, Bradley and Juravsky, Jordan and Ehrlich, Ryan and Clark, Ronald and Le, Quoc V. and Ré, Christopher and Mirhoseini, Azalia},
  date = {2024-12-30},
  eprint = {2407.21787},
  eprinttype = {arXiv},
  eprintclass = {cs},
  doi = {10.48550/arXiv.2407.21787},
  url = {http://arxiv.org/abs/2407.21787},
  urldate = {2026-02-11},
  pubstate = {prepublished}
}

@online{alphaevolve,
  title = {{{AlphaEvolve}}: {{A}} Coding Agent for Scientific and Algorithmic Discovery},
  shorttitle = {{{AlphaEvolve}}},
  author = {Novikov, Alexander and Vũ, Ngân and Eisenberger, Marvin and Dupont, Emilien and Huang, Po-Sen and Wagner, Adam Zsolt and Shirobokov, Sergey and Kozlovskii, Borislav and Ruiz, Francisco J. R. and Mehrabian, Abbas and Kumar, M. Pawan and See, Abigail and Chaudhuri, Swarat and Holland, George and Davies, Alex and Nowozin, Sebastian and Kohli, Pushmeet and Balog, Matej},
  date = {2025-06-16},
  eprint = {2506.13131},
  eprinttype = {arXiv},
  eprintclass = {cs},
  doi = {10.48550/arXiv.2506.13131},
  url = {http://arxiv.org/abs/2506.13131},
  urldate = {2026-02-11},
  pubstate = {prepublished}
}

@misc{polybench,
    title = {{PolyBench}},
    url = {https://web.cse.ohio-state.edu/~pouchet.2/software/polybench/},
    urldate = {2022-07-24},
    author = {Pouchet, Louis-Noël and Bondugula, Uday},
}

@inproceedings{machsuite,
    address = {Raleigh, North Carolina},
    title = {{MachSuite}: {Benchmarks} for {Accelerator} {Design} and {Customized} {Architectures}},
    booktitle = {Proceedings of the {IEEE} {International} {Symposium} on {Workload} {Characterization}},
    author = {Reagen, Brandon and Adolf, Robert and Shao, Yakun Sophia and Wei, Gu-Yeon and Brooks, David},
    month = oct,
    year = {2014},
}

@inproceedings{chstone,
  title = {{{CHStone}}: {{A}} Benchmark Program Suite for Practical {{C-based}} High-Level Synthesis},
  shorttitle = {{{CHStone}}},
  booktitle = {2008 {{IEEE International Symposium}} on {{Circuits}} and {{Systems}} ({{ISCAS}})},
  author = {Hara, Yuko and Tomiyama, Hiroyuki and Honda, Shinya and Takada, Hiroaki and Ishii, Katsuya},
  date = {2008-05},
  pages = {1192--1195},
  issn = {2158-1525},
  doi = {10.1109/ISCAS.2008.4541637},
  eventtitle = {2008 {{IEEE International Symposium}} on {{Circuits}} and {{Systems}} ({{ISCAS}})}
}

@inproceedings{rosetta,
  title = {Rosetta: {{A Realistic High-Level Synthesis Benchmark Suite}} for {{Software Programmable FPGAs}}},
  shorttitle = {Rosetta},
  booktitle = {Proceedings of the 2018 {{ACM}}/{{SIGDA International Symposium}} on {{Field-Programmable Gate Arrays}}},
  author = {Zhou, Yuan and Gupta, Udit and Dai, Steve and Zhao, Ritchie and Srivastava, Nitish and Jin, Hanchen and Featherston, Joseph and Lai, Yi-Hsiang and Liu, Gai and Velasquez, Gustavo Angarita and Wang, Wenping and Zhang, Zhiru},
  date = {2018-02-15},
  series = {{{FPGA}} '18},
  pages = {269--278},
  publisher = {Association for Computing Machinery},
  location = {New York, NY, USA},
  doi = {10.1145/3174243.3174255},
  url = {https://dl.acm.org/doi/10.1145/3174243.3174255},
  urldate = {2025-10-15},
  isbn = {978-1-4503-5614-5}
}

@article{c2hlscb,
  title = {{{C2HLSC}}: {{Leveraging Large Language Models}} to {{Bridge}} the {{Software-to-Hardware Design Gap}}},
  shorttitle = {{{C2HLSC}}},
  author = {Collini, Luca and Garg, Siddharth and Karri, Ramesh},
  date = {2025-05-10},
  journaltitle = {ACM Trans. Des. Autom. Electron. Syst.},
  issn = {1084-4309},
  doi = {10.1145/3734524},
  url = {https://dl.acm.org/doi/10.1145/3734524},
  urldate = {2025-10-16}
}

@article{democratizing_dsc,
  title = {Democratizing {{Domain-Specific Computing}}},
  author = {Chi, Yuze and Qiao, Weikang and Sohrabizadeh, Atefeh and Wang, Jie and Cong, Jason},
  date = {2022-12-20},
  journaltitle = {Commun. ACM},
  volume = {66},
  number = {1},
  pages = {74--85},
  issn = {0001-0782},
  doi = {10.1145/3524108},
  url = {https://dl.acm.org/doi/10.1145/3524108},
  urldate = {2025-02-28}
}

@online{licht2020transformations,
  title = {Transformations of {{High-Level Synthesis Codes}} for {{High-Performance Computing}}},
  author = {Licht, Johannes de Fine and Besta, Maciej and Meierhans, Simon and Hoefler, Torsten},
  date = {2020-11-23},
  eprint = {1805.08288},
  eprinttype = {arXiv},
  eprintclass = {cs},
  doi = {10.48550/arXiv.1805.08288},
  url = {http://arxiv.org/abs/1805.08288},
  urldate = {2025-10-16},
  pubstate = {prepublished}
}

@online{llm_code,
  title = {Evaluating {{Large Language Models Trained}} on {{Code}}},
  author = {Chen, Mark and Tworek, Jerry and Jun, Heewoo and Yuan, Qiming and Pinto, Henrique Ponde de Oliveira and Kaplan, Jared and Edwards, Harri and Burda, Yuri and Joseph, Nicholas and Brockman, Greg and Ray, Alex and Puri, Raul and Krueger, Gretchen and Petrov, Michael and Khlaaf, Heidy and Sastry, Girish and Mishkin, Pamela and Chan, Brooke and Gray, Scott and Ryder, Nick and Pavlov, Mikhail and Power, Alethea and Kaiser, Lukasz and Bavarian, Mohammad and Winter, Clemens and Tillet, Philippe and Such, Felipe Petroski and Cummings, Dave and Plappert, Matthias and Chantzis, Fotios and Barnes, Elizabeth and Herbert-Voss, Ariel and Guss, William Hebgen and Nichol, Alex and Paino, Alex and Tezak, Nikolas and Tang, Jie and Babuschkin, Igor and Balaji, Suchir and Jain, Shantanu and Saunders, William and Hesse, Christopher and Carr, Andrew N. and Leike, Jan and Achiam, Josh and Misra, Vedant and Morikawa, Evan and Radford, Alec and Knight, Matthew and Brundage, Miles and Murati, Mira and Mayer, Katie and Welinder, Peter and McGrew, Bob and Amodei, Dario and McCandlish, Sam and Sutskever, Ilya and Zaremba, Wojciech},
  date = {2021-07-14},
  eprint = {2107.03374},
  eprinttype = {arXiv},
  eprintclass = {cs},
  doi = {10.48550/arXiv.2107.03374},
  urldate = {2025-02-28},
  pubstate = {prepublished}
}

@inproceedings{hdleval,
  title = {{{HDLEval Benchmarking LLMs}} for Multiple {{HDLs}}},
  booktitle = {2024 {{IEEE LLM Aided Design Workshop}} ({{LAD}})},
  author = {Kashanaki, Farzaneh Rabiei and Zakharov, Mark and Renau, Jose},
  date = {2024-06},
  pages = {1--5},
  doi = {10.1109/LAD62341.2024.10691770},
  urldate = {2025-02-28},
  eventtitle = {2024 {{IEEE LLM Aided Design Workshop}} ({{LAD}})}
}

@online{verilog_tool_feedback,
  title = {Can {{EDA Tool Feedback Improve Verilog Generation}} by {{LLMs}}?},
  author = {Blocklove, Jason and Thakur, Shailja and Tan, Benjamin and Pearce, Hammond and Garg, Siddharth and Karri, Ramesh},
  date = {2024-11-01},
  eprint = {2411.11856},
  eprinttype = {arXiv},
  eprintclass = {cs},
  doi = {10.48550/arXiv.2411.11856},
  urldate = {2025-02-28},
  pubstate = {prepublished}
}

@inproceedings{verilogeval,
  title = {{{VerilogEval}}: Evaluating Large Language Models for Verilog Code Generation},
  booktitle = {2023 {{IEEE}}/{{ACM}} International Conference on Computer-Aided Design ({{ICCAD}})},
  author = {Liu, Mingjie and Pinckney, Nathaniel and Khailany, Brucek and Ren, Haoxing},
  date = {2023}
}

@misc{verilogeval_v2,
  title = {Revisiting {{VerilogEval}}: {{Newer}} Llms, in-Context Learning, and Specification-to-{{RTL}} Tasks},
  author = {Pinckney, Nathaniel and Batten, Christopher and Liu, Mingjie and Ren, Haoxing and Khailany, Brucek},
  date = {2024},
  eprint = {2408.11053},
  eprinttype = {arXiv},
  eprintclass = {cs.SE},
}

@inproceedings{scalehls,
  title = {{{ScaleHLS}}: {{A New Scalable High-Level Synthesis Framework}} on {{Multi-Level Intermediate Representation}}},
  shorttitle = {{{ScaleHLS}}},
  booktitle = {2022 {{IEEE International Symposium}} on {{High-Performance Computer Architecture}} ({{HPCA}})},
  author = {Ye, Hanchen and Hao, Cong and Cheng, Jianyi and Jeong, Hyunmin and Huang, Jack and Neuendorffer, Stephen and Chen, Deming},
  date = {2022-04},
  pages = {741--755},
  issn = {2378-203X},
  doi = {10.1109/HPCA53966.2022.00060},
  url = {https://ieeexplore.ieee.org/abstract/document/9773203},
  urldate = {2025-10-16},
  eventtitle = {2022 {{IEEE International Symposium}} on {{High-Performance Computer Architecture}} ({{HPCA}})}
}

@article{allo,
  title = {Allo: {{A Programming Model}} for {{Composable Accelerator Design}}},
  shorttitle = {Allo},
  author = {Chen, Hongzheng and Zhang, Niansong and Xiang, Shaojie and Zeng, Zhichen and Dai, Mengjia and Zhang, Zhiru},
  date = {2024-06-20},
  journaltitle = {Allo: A Programming Model for Composable Accelerator Design},
  shortjournal = {Proc. ACM Program. Lang.},
  volume = {8},
  pages = {171:593--171:620},
  doi = {10.1145/3656401},
  url = {https://dl.acm.org/doi/10.1145/3656401},
  urldate = {2025-10-16},
  issue = {PLDI}
}

@online{dato,
  title = {Dato: {{A Task-Based Programming Model}} for {{Dataflow Accelerators}}},
  shorttitle = {Dato},
  author = {Fang, Shihan and Chen, Hongzheng and Zhang, Niansong and Li, Jiajie and Meng, Han and Liu, Adrian and Zhang, Zhiru},
  date = {2025-09-08},
  eprint = {2509.06794},
  eprinttype = {arXiv},
  eprintclass = {cs},
  doi = {10.48550/arXiv.2509.06794},
  url = {http://arxiv.org/abs/2509.06794},
  urldate = {2025-10-16},
  pubstate = {prepublished}
}

@online{kvapil2025intelligent,
  title = {Intelligent Experiments through Real-Time {{AI}}: {{Fast Data Processing}} and {{Autonomous Detector Control}} for {{sPHENIX}} and Future {{EIC}} Detectors},
  shorttitle = {Intelligent Experiments through Real-Time {{AI}}},
  author = {Kvapil, J. and Borca-Tasciuc, G. and Bossi, H. and Chen, K. and Chen, Y. and Morales, Y. Corrales and Costa, H. Da and Silva, C. Da and Dean, C. and Durham, J. and Fu, S. and Hao, C. and Harris, P. and Hen, O. and Jheng, H. and Lee, Y. and Li, P. and Li, X. and Lin, Y. and Liu, M. X. and Loncar, V. and Mitrevski, J. P. and Olvera, A. and Purschke, M. L. and Renck, J. S. and Roland, G. and Schambach, J. and Shi, Z. and Tran, N. and Wuerfel, N. and Xu, B. and Yu, D. and Zhang, H.},
  date = {2025-01-08},
  eprint = {2501.04845},
  eprinttype = {arXiv},
  eprintclass = {physics},
  doi = {10.48550/arXiv.2501.04845},
  url = {http://arxiv.org/abs/2501.04845},
  urldate = {2025-10-16},
  pubstate = {prepublished}
}

@article{Wan2022RoboticCO,
  title = {Robotic Computing on Fpgas: {{Current}} Progress, Research Challenges, and Opportunities},
  author = {Wan, Zishen and Lele, Ashwin Sanjay and Yu, Bo and Liu, Shaoshan and Wang, Yu and Reddi, Vijay Janapa and Hao, Cong and Raychowdhury, Arijit},
  date = {2022},
  journaltitle = {2022 IEEE 4th International Conference on Artificial Intelligence Circuits and Systems (AICAS)},
  pages = {291--295},
  url = {https://api.semanticscholar.org/CorpusID:248811737}
}

@article{duarte2018fasta,
  title = {Fast Inference of Deep Neural Networks in {{FPGAs}} for Particle Physics},
  author = {Duarte, Javier and Han, Song and Harris, Philip and Jindariani, Sergo and Kreinar, Edward and Kreis, Benjamin and Ngadiuba, Jennifer and Pierini, Maurizio and Rivera, Ryan and Tran, Nhan and Wu, Zhenbin},
  date = {2018-07-27},
  journaltitle = {Journal of Instrumentation},
  shortjournal = {J. Inst.},
  volume = {13},
  number = {07},
  eprint = {1804.06913},
  eprinttype = {arXiv},
  eprintclass = {physics},
  pages = {P07027-P07027},
  issn = {1748-0221},
  doi = {10.1088/1748-0221/13/07/P07027},
  url = {http://arxiv.org/abs/1804.06913},
  urldate = {2025-10-16}
}

@article{chen2024understanding,
  title = {Understanding the {{Potential}} of {{FPGA-based Spatial Acceleration}} for {{Large Language Model Inference}}},
  author = {Chen, Hongzheng and Zhang, Jiahao and Du, Yixiao and Xiang, Shaojie and Yue, Zichao and Zhang, Niansong and Cai, Yaohui and Zhang, Zhiru},
  date = {2024-12-17},
  journaltitle = {ACM Trans. Reconfigurable Technol. Syst.},
  volume = {18},
  number = {1},
  pages = {5:1--5:29},
  issn = {1936-7406},
  doi = {10.1145/3656177},
  url = {https://dl.acm.org/doi/10.1145/3656177},
  urldate = {2025-10-16}
}

\end{document}